\documentclass[11pt, graphicx]{article}
\usepackage{delarray}
\usepackage{epsfig}

\begin{document} 

\begin{center}
\Large\textbf {Particle Energy And Acceleration Efficiencies
 In Highly Relativistic Shocks\footnote{Results presented in the \textbf
{\textit{17th ECRS}}, July 2000, Lodz Poland}}
\end{center}

\begin{center}
\renewcommand{\thefootnote}{\fnsymbol{footnote}} 
\Large A.Meli and J.J.Quenby
\end{center}
\begin{center}
{\Large{\textit {Astrophysics Group, Blackett Laboratory}}}
{\Large{\textit{Imperial College of Science,Technology and Medicine, London, 
UK}}}
\end{center}

\begin{abstract}
{\small\textbf{In this paper we present an investigation of  
 numerical Monte Carlo  simulations of the diffusive shock acceleration 
 in the test particle limit. Very high gamma flow astrophysical plasmas, 
 have been used, from $\gamma_{up}$ $\sim50$ up to $\gamma_{up}$ $\sim1000$
,which could be relevant to the suggested models 
of AGNs Jets and their Central Engines as well as the ultra-relativistic 
shock particle 
acceleration in Gamma Ray 
Burst (GRB) fireballs. Particularly the energy  gain per shock crossing 
and the 
time constant  for the above high relativistic processes is numerically 
calculated. We explicitly find a considerable $\gamma^{2}$ energy boosting 
in the first shock cycle, and in all subsequent shock cycles the particle energy is multiplied by a large factor. Also a noted acceleration 
speed-up for the same acceleration process has been observed. Both of 
those results are connected with theoretical suggestions that a significant enhancement of the acceleration process is possible due to relativistic effects not present at  
lower plasma flow speeds. The acceleration speed-up and the mean energy-gain per shock cycle 
found could efficiently justify the theory of the 
origin of Ultra High Energy Cosmic Rays (UHECR) from the sites of GRBs.}}
\end{abstract}

\section{\large{Introduction}}

{\large {Ultra-relativistic flow velocities are found in many 
astrophysical objects such as in Gamma Ray Bursts (GRB) as well as in AGN jets 
and their central engines where accretion is observed. 
The current understanding of multiwavelenght GRB afterglow observations, 
indicates that they are produced at the relativistic shocks, formed
 shortly after a violent explosion takes place where a blast wave impacts 
with the surrounding interstellar medium. Consequently shock waves form (Meszaros \& Rees, 1993), and eventually relativistic diffusive shock 
acceleration takes place.
Theoretical works of Vietri (1995,1997 and 1998) and Waxman (1995) suggested 
that diffusive acceleration at non-relativistic shocks is not 
fast enough to produce the highest energy particles observed, 
but they considered that a potential relativistic supesonic environment could 
succeed to do so. One of the first computational works 
(Quenby \& Lieu, 1989)) concerning relativistic ($\gamma\sim$3)  
diffusive acceleration with application to AGN relativistic flows  
reported a considerable enhancement in the acceleration rate of a 
factor$\sim$13.
In this present work we present a numerical investigation for diffusive 
acceleration in ultra-relativistic shocks with application to GRBs 
theoretical models.

The standard analytical theory of diffusive shock acceleration 
operating in one dimensional astrophysical non-relativistic flows, 
where $u_{1}$, $u_{2}$ are the non-relativistic upstream and downstream flow 
speeds and the particle's velocity is almost equal to c, yields a differential 
spectrum  $dn/dp \propto p^{-\alpha}$, where $\alpha=r+2/r+1$ and 
$r=u_{1}/u_{2}$ and the analytically
derived acceleration time constant is given by, 

\begin{equation}
t_{acc}=\frac{c}{u_{1}-u_{2}}[\frac{\lambda_{1}}{u_{1}}+
\frac{\lambda_{2}}{u_{2}}]\\
\end{equation}\\
where $\lambda_{1}$, $\lambda_{2}$  are the particle's mean free paths, 
upstream and downstream respectively. It is known that the fractional energy 
increase 
for a single shock crossing, for relativistic flows, 
is $\sim (1-V^{2}/c^{2})^{-0.5})$ where V is the relativistic difference 
($u_{1}-u_{2}$) in flow velocities across the shock. Considering again 
relativistic flows and diffusion theory calculations for
 the cycle-time constant, it is found that the latter does not change from the non-relativistic case, and so we could expect a speed-up of 
acceleration,  relative to the non-relativistic estimate. 
Because relativistic flows create large anisotropies
in the plasma flow a detailed computational approach is necessary
in this high $\gamma$ regime. Several studies have been made by many authors, 
but  still, there is a controversy about the rate of particles energy 
gain and the time constant of the acceleration. These controversies   
need a solution as they are crucial for models of AGN jet 
cosmic ray acceleration and UHECR production and acceleration in GRBs 
(eg Vietri). In this present work we present a 
Monte Carlo simulation, investigating the diffusive 
acceleration in ultra-relativistic parallel shocks for values up to 
$\gamma\sim990$, with application to GRBs relativistic flows.
We find a considerable acceleration rate enhancement and our results 
show explicitly that the energy boosting, scales as $\gamma^{2}$ in 
the first shock passaging cycle. After that, the particle energy is 
multiplied by a large number, gaining large amounts 
of energy which could still account for the theory of UHECR. These   
results support theoretical predictions (eg. Vietri 1995)
 concerning UHECR origin and acceleration from GRBs' highly relativistic plasma flows.

\section{\large{Numerical method}}

In this Monte Carlo code we consider both isotropic large angle and 
pitch angle diffusion which is calculated in the respective 
plasma rest frames, where  particles are allowed  to scatter in the respective fluid frames towards the shock. We consider only parallel shock ($\theta=0$, where $\theta$ is the angle between the shock normal and the magnetic field), 
either because that is the field configuration,  or turbulence removes 
"reflection" at the interface. A guiding centre approximation is used  
where the particle trajectory is followed in one-dimensional space along the 
x axis. A relativistic transformation is performed to the local plasma frames 
each time the particle scatters across the shock 
following it according to particle \textit{jump conditions} and it
is made leave the system from the moment that it 'escapes'  far downstream 
at the spatial boundary or if it reaches a well defined maximum energy.
The particles ($\sim$$10^{7}$) of weight equal to one, are injected far 
upstream at a constant 
energy of high gamma, which supposes that a pre-acceleration of the particles 
has already taken  place. They left to move towards the shock where along the way they collide with the pressumed scattering centers and consequently as 
they keep scattering  between the upstream and downstream regions they gain 
each time an amount of energy. The compresion ratio is allowed to have the value of 4, for immediate comparison with the non-relativistic values. 
A main characteristic of the code is that a particle \textit{splitting 
technique} 
is used, in order for  the statistics to be as efficient as possible. 
The main notion of particle splitting is the fact 
that one can consider that each particle represents instead  
a large number of particles. When a
number of particles escapes through the defined spatial or momentum 
boundaries during the acceleration process, we replace these \textit{lost} particles with new ones -in order to have again the same initial number of particles- but in such a way that as their weight is decreased, the number of particles remains almost constant throughout the simulation.
The probability that a particle will move a distance z 
along the field lines at pitch angle $\theta$ before it scatters
is given by the following expression,

\begin{equation}
Prob(z) \sim exp(-z/\lambda|cos\theta|)
\end{equation}\\
For the case of  pitch angle diffusion scattering, the new pitch angle 
$\theta'$ is calculated by the trigonometric formula,

\begin{equation}
cos\theta'=cos\theta\sqrt{1-sin^2\delta\theta}+sin\delta\theta\sqrt{1-
cos^2\theta} cos\phi
\end{equation}\\
where $\phi\epsilon$(0,$2\pi$) is the azimuth angle with respect to 
the original 
momentum direction. We note here that for the case of highly relativistic 
flows, the definition of particles pitch angle diffusion is 
scattering within an angle $\sim$ 1/$\gamma_{1}$, where $\gamma_{1}$ is
the upstream gamma measured in the shock frame (Gallant \& Achterberg, 1999).
This important condition is explicitly 
included in our Monte Carlo simulations during the 
relativistic particle shock acceleration process, 
having as a main aim  to investigate and compare our results with  
controversial theoretical predictions which depend whether either large angle or pitch angle diffusion operates in highly relativistic flows.

\section{\large{Results}}
\begin{figure}
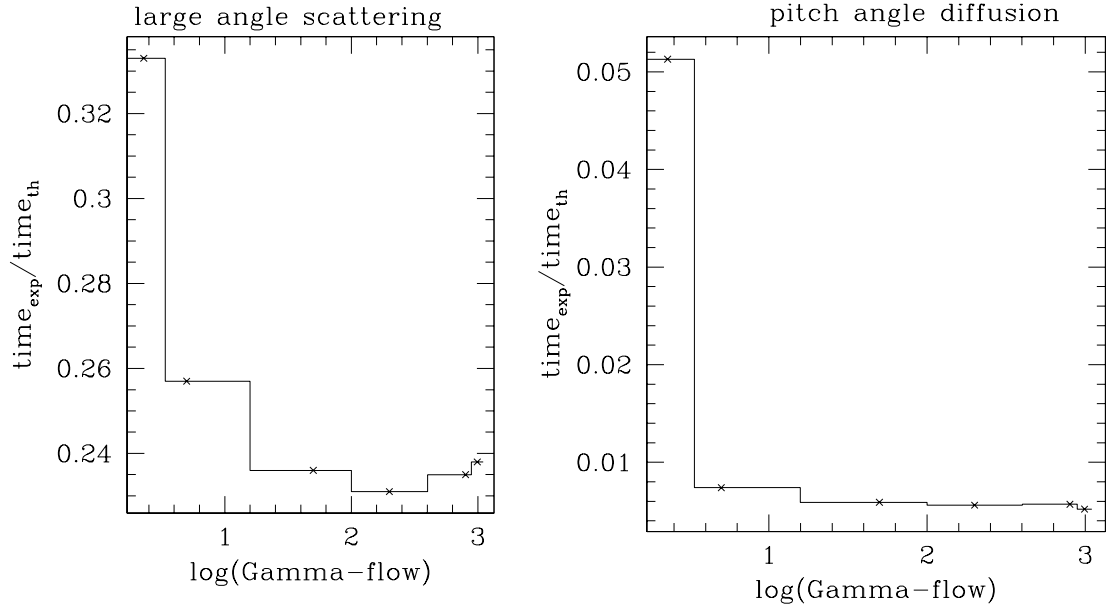

\begin{minipage}{170mm}
\epsfig{figure=plot.lg.time,width=80mm}
\epsfig{figure=plot.sm2.time,width=80mm}
\end{minipage}
\caption{\textit{Left}--The ratio of the computational to the non-relativistic
analytical time constants for large angle scattering versus the  
Lorentz $\gamma_{up}$ flow.\textit{Right}--The equivalent plot for pitch angle 
diffusion. As it is seen, for both plots there is a maximum substantial 
'speed up' of a factor $\sim$5 and $\sim$200, respectively.}
\end{figure}

\begin{figure}
\begin{minipage}{170mm}
\epsfig{figure=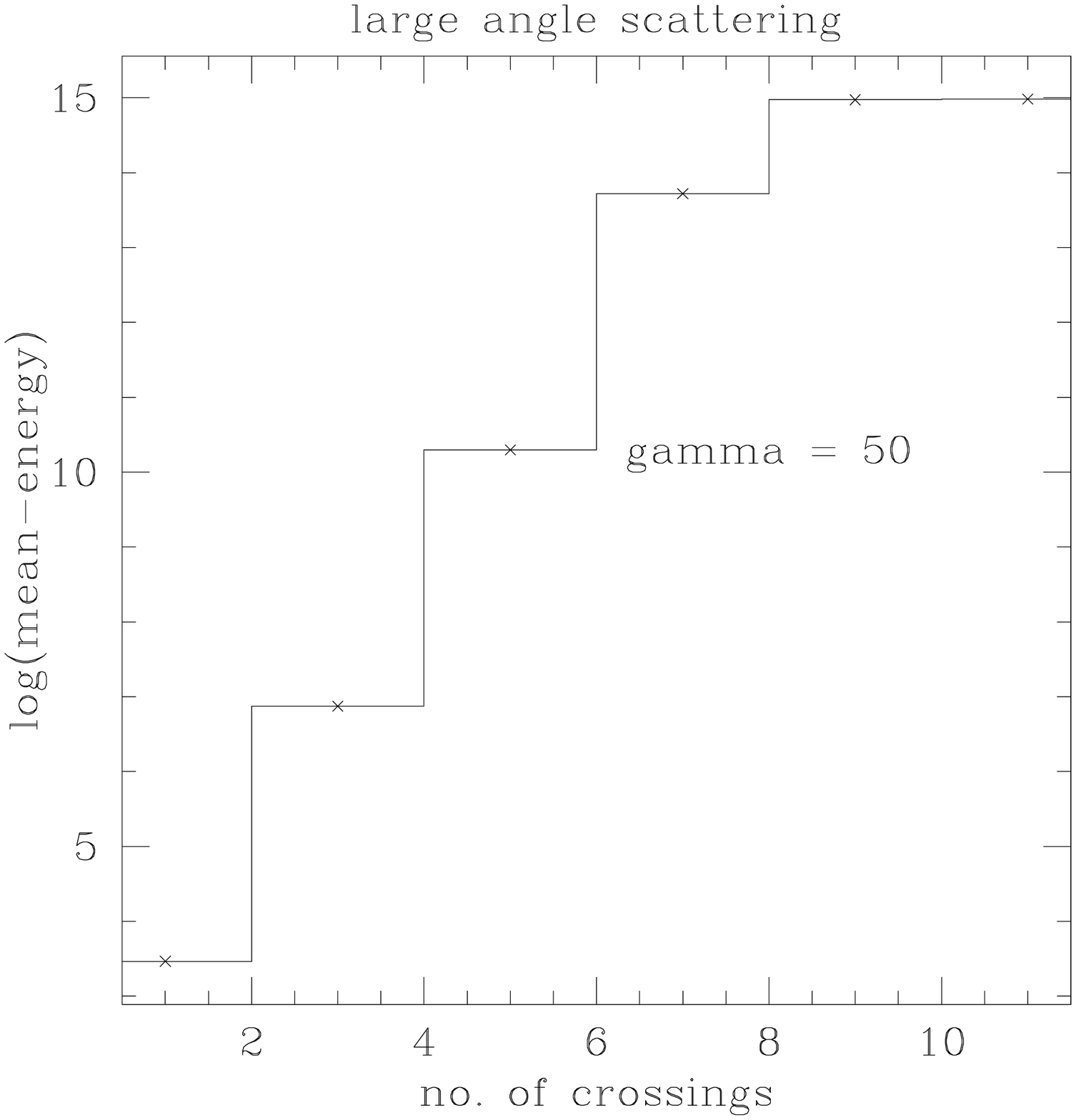,width=80mm}
\epsfig{figure=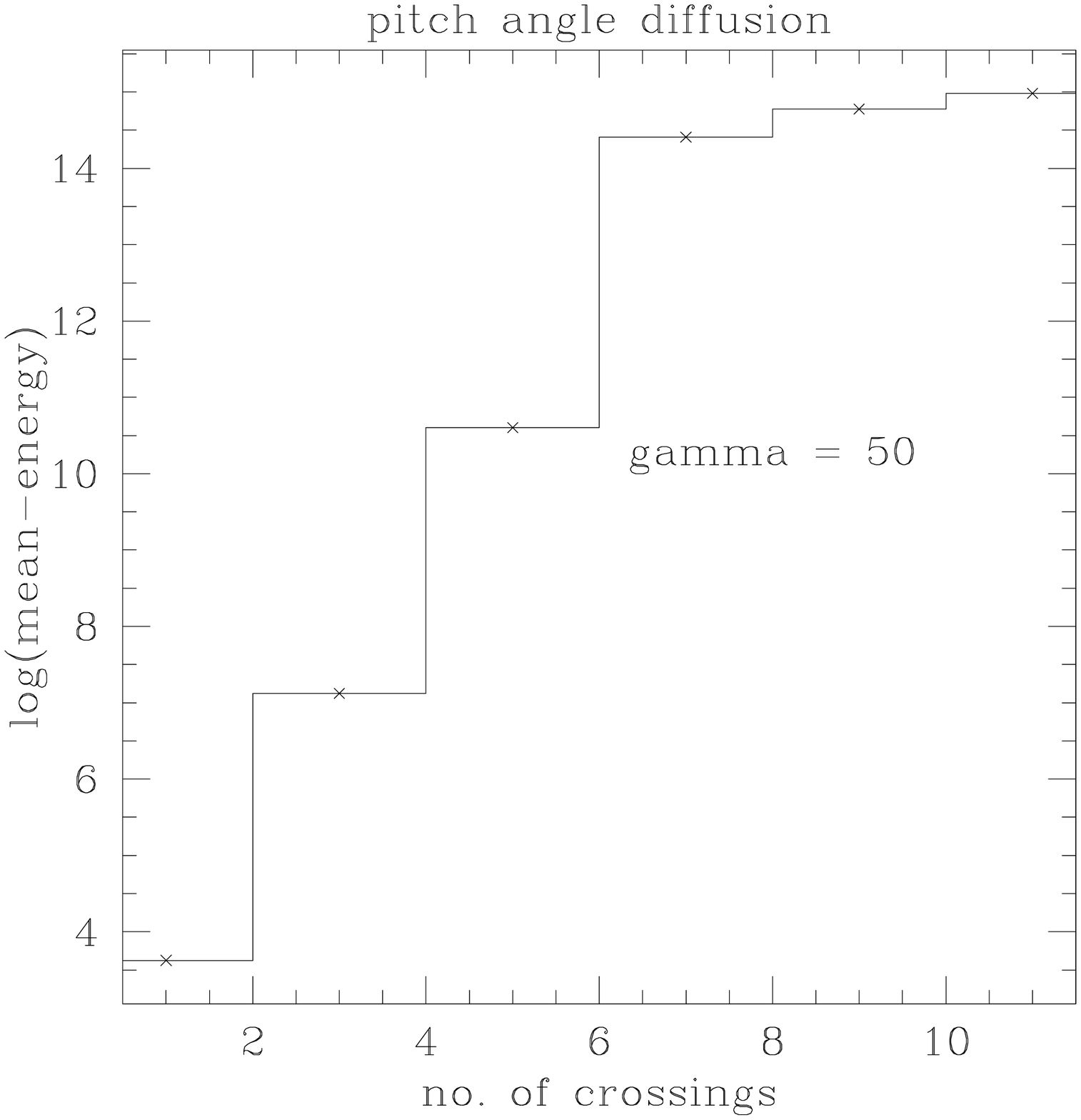,width=80mm}
\caption{\textit{Left}: For large angle scattering we calculate the mean energy of the 
particle versus the no. of shock crossings, employing
an upstream flow gamma factor $\gamma$=50. We may observe the efficiency of the $\gamma^{2}$
boosting in the first cycle and the large energy multiplication in all subsequent crossings.
\textit{Right}: For pitch angle diffusion the particle energy gain appears (normally) to be less high.}
\epsfig{figure=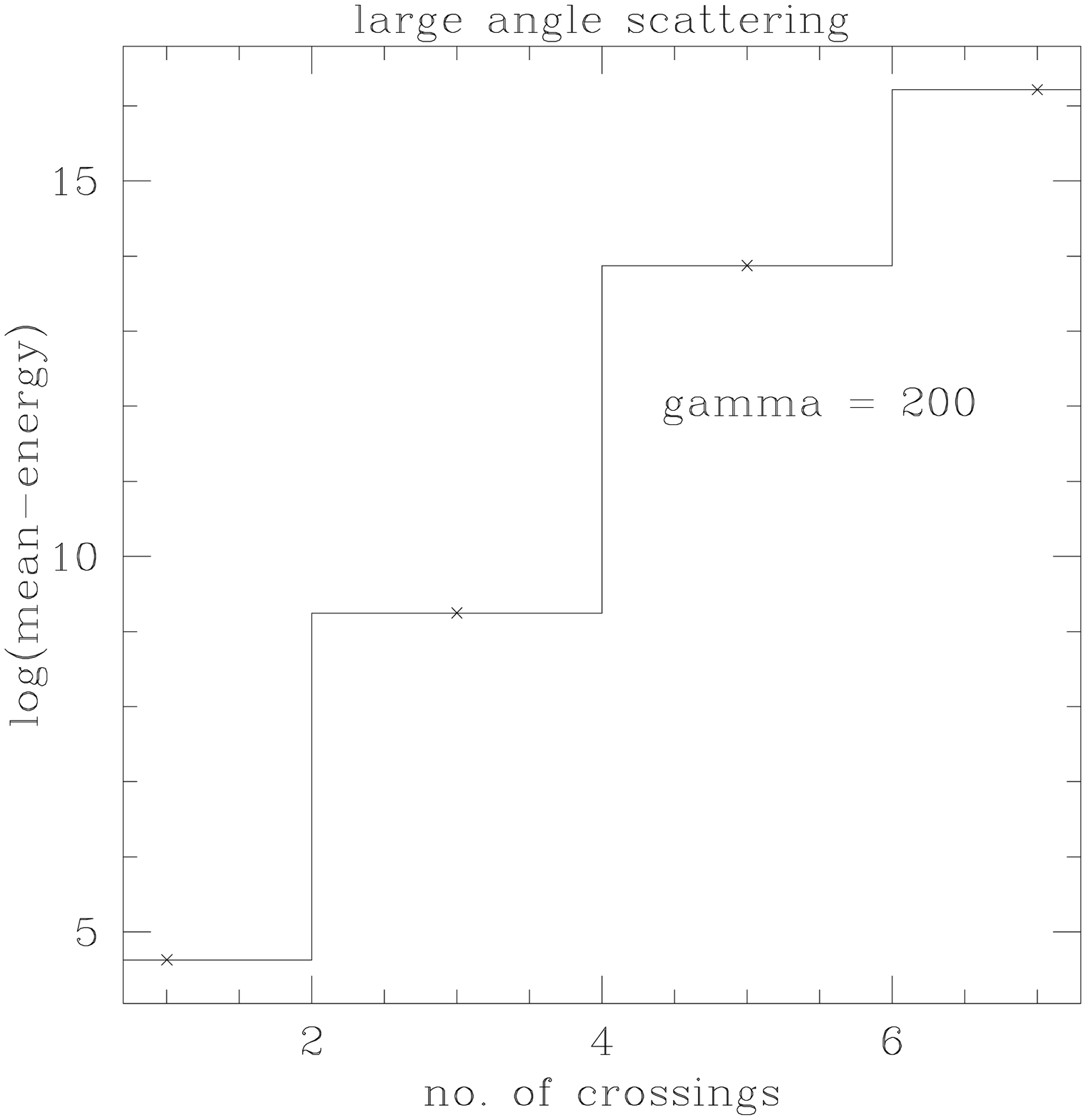,width=80mm}
\epsfig{figure=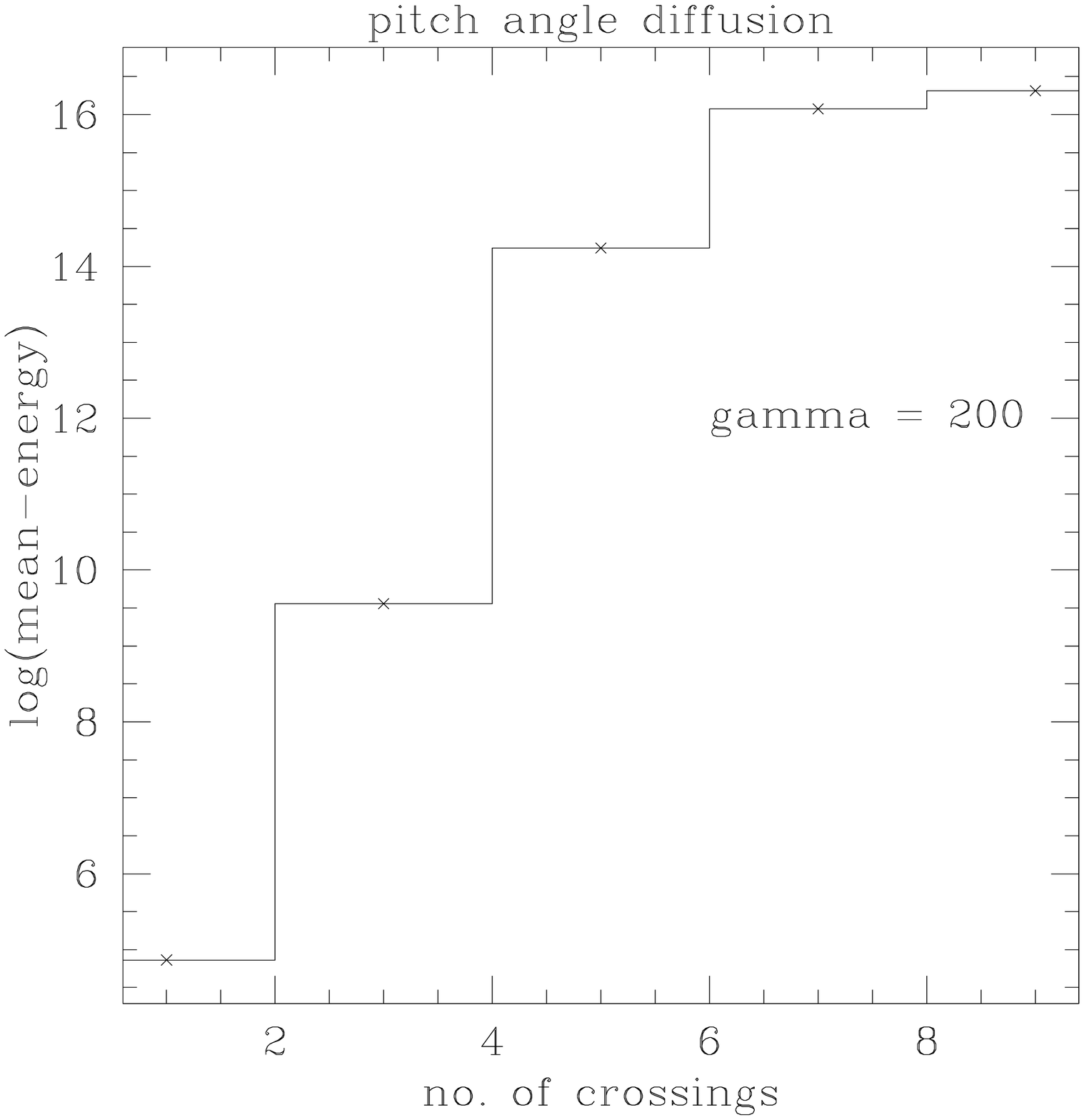,width=80mm}
\caption{\textit{Left}: In this plot we use large angle diffusion 
and the upstream flow has the value of $\gamma$=200. It is seen again the 
efficiency of the mean energy gain in every shock crossing
\textit{Right}: For pitch angle diffusion, the mean energy gain per cycle for 
$\gamma$=200 versus the shock crossing, gives a considerable energy gain
efficiency mostly in the first shock cycle crossing ($\gamma^{2}$).}
\end{minipage}
\end{figure}

\begin{figure}
\epsfig{figure=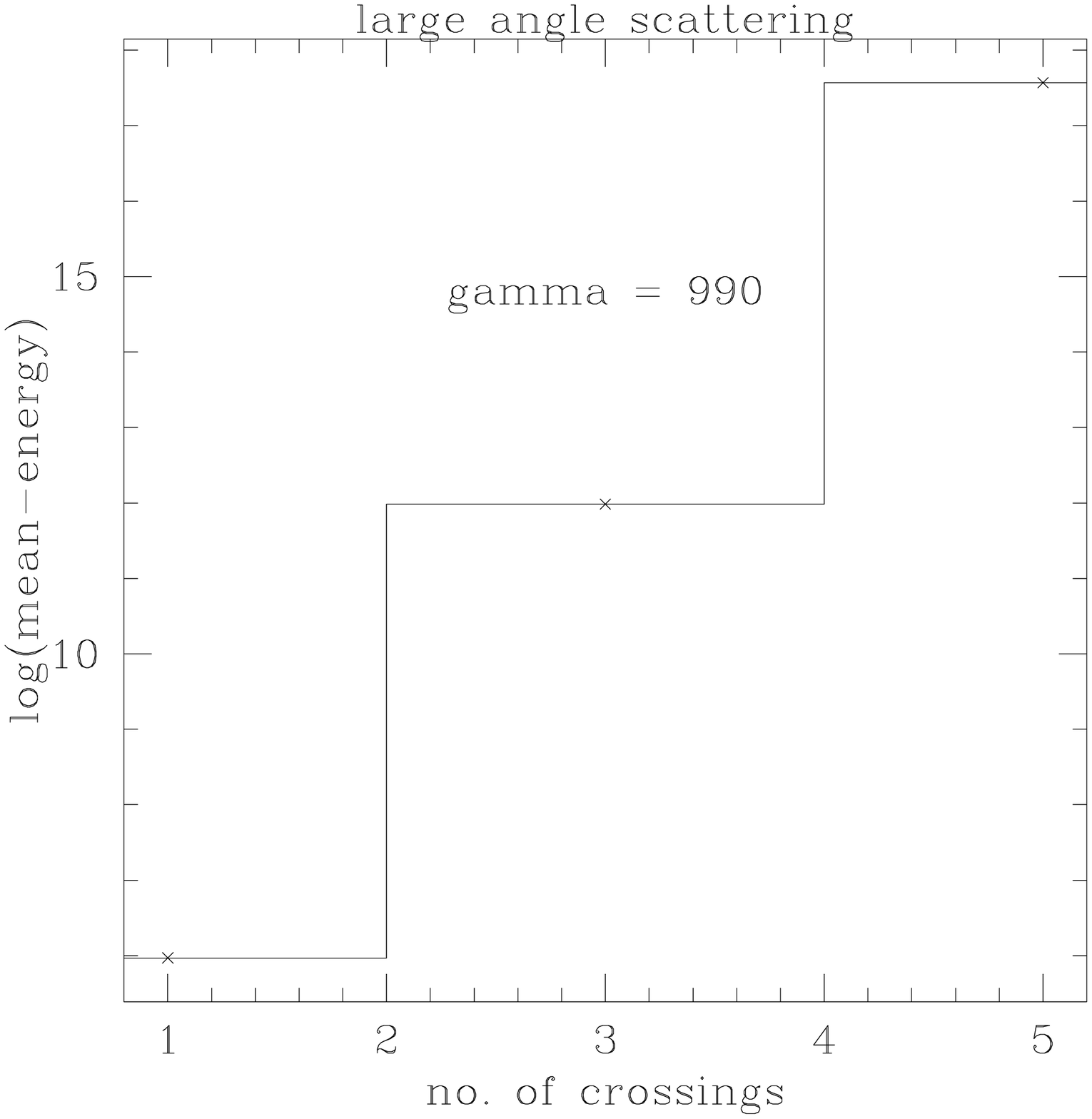,width=92mm}
\caption{In this plot we use $\gamma$=990 for large angle diffusion.
One may observe that the $\gamma^{2}$ energy boosting efficiency is explicitly present  
in the first  shock cycle.}
\epsfig{figure= 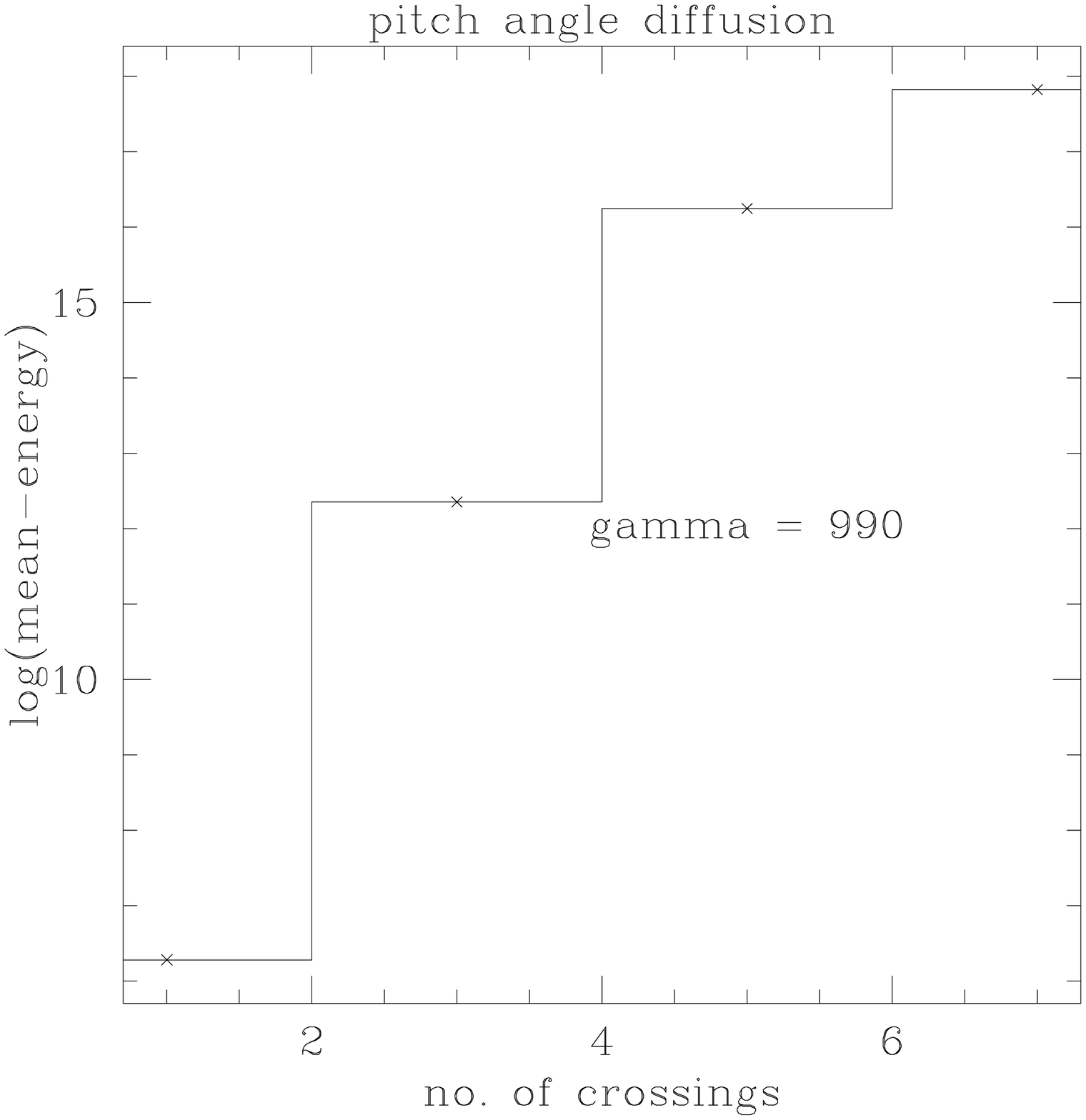,width=92mm}
\caption{For pitch angle diffusion, the mean energy gain per cycle 
for $\gamma$=990, to be compared  with the square of the upstream 
flow gamma factor $\gamma^{2}$ for the first shock cycle.}
\end{figure}

The aim of this work is to investigate the energy boosting of particles 
accelerated in higly relativistic shock waves in comparison to the shock 
cycle and to study the cycle time constant 
for the particle acceleration, by simulating highly relativistic shock
 formations 
which potentially occur in GRBs. 
The results shown in the figures, have been recorded in 
the downstream side, in the shock  frame and concern  large angle 
isotropic scattering and  pitch angle diffusion  for a various number of  
upstream Lorentz $\gamma$ values. Figure 1, shows for both large angle 
scattering and pitch angle diffusion, 
the  ratio of the computational time constant to the non-relativistic 
analytical acceleration time constant, as a function of the upstream Lorentz 
factor
$\gamma_{up}$ plasma flows. We note here that in the case of the pitch angle  diffusion, we allow for 
the mean free path of particle to scatter 
$\sim \pi/2$ for the calculation of the analytical time-expression. 
We can observe that substantial 'speed-up' of a  factor $\sim$ 5 for 
large angle scattering  is given and a considerable 
'speed-up' of  a factor $\sim$ 200 is noted for pitch angle diffusion.
The figures 2-5 show the mean energy gain of the accelerated particles 
versus the shock crossing where a substantial $\gamma^{2}$ boosting is
found, explicitly for the first cycle. After that, and for the following 
shock crossings, the energy of the particle is multiplied by a large factor. 
Our first indication ($\gamma^{2}$) is in good aggreement with Gallant \& Achterberg (1999)
predictions who found \textit{'that the energy of the particle for the first shock
cycle is  increased by a factor of $\gamma_{up}^{2}$'.} 
On the other hand though, they found that \textit{...'in all subsequent shock crossing 
cycles the particle energy doubles'} and that due to the
fact that the particle does not have time to re-isotropise upstream before being overtaken 
by the shock. Our results show that the particle energy is multiplied by a large factor. 
This could be justified from the moment we could assume that the acceleration speed-up found,
\textit{gives} enough time for the particle to re-isotropise upstream before being 
overtaken by the shock.

\section{\large{Conclusions}}

A Monte Carlo numerical investigation has been reported in the 
test particle limit of parallel diffusive shock acceleration. 
Very high gamma flow astrophysical plasmas have been used, 
from $\gamma_{up}$ $\sim50$  up to $\gamma_{up}$  $\sim1000$ which are relevant to 
highly relativistic astrophysical supersonic 
plasma flows in GRBs. A dramatic  $\gamma^{2}$ energy boost in the first shock crossing 
cycle has 
been observed and a considerable acceleration speed-up has 
been found for both isotropic large angle and pitch angle diffusion. 
Both these results support Vietri and Waxman GRB models and theoretical 
predictions for the model of Ultra High Energy Cosmic Ray origin from GRBs. 
The surprising implication of those results is that a modeler could consider
that, indeed there is a distinctive case 
showing that a $\gamma^{2}$ energy gain versus first shock crossing and a large
energy multiplication in all other subsequent cycles is 
indeed happening in highly relativistic shock waves, 'making space' to the particles 
in order to gain rather very large amounts of energy in few shock cycles. 
On the other hand the 
above implications, along with our numerical calculations, finding a considerable 
acceleration speed-up could sufficiently account for the theoretical 
predictions implying UHECR origin from GRB relativistic shock configurations.
However these crucial issues, need further detailed investigation as many 
other parameters need to be included within the simulation codes such as shock obliquity, 
non-linear processes and energy losses in order for potential models to be as realistic 
as possible with the actual relativistic flow of GRBs.  There is already under way a 
similar work concerning the above parameters.}}
\\
\\
{\large{\textbf{Acknowledgments}}: A.Meli wishes to thank 
J.Kirk and M.Ostrowki for valuable discussions during the 
$\textit {$17^{th}$ECRC 2000}$ in Lodz, Poland.

\section{\large{\textit{References}}}
{\small
Baring, M.G, ICRC Salt Lake City, Proceedings (1999)\\
Bednarz, J. \& Ostrowski, M. MNRAS 283, 447-456(1996)\\
Bednarz J. MNRAS 315L.37B(2000)\\
Bell, A.R., MNRAS 182,147-156(1978)\\
Bell, A.R , MNRAS 182,443-455(1978)\\
Blanfdord, R.D $\&$ Ostriker, J.P ApJ 211, 793-808(1977)\\
Ellison, D.C.,\& Jones, F.C., \& Reynolds,S.P., ApJ 360, 702 (1990a)\\
Gallant, Y.A., \& Achterberg A., MNRAS 305L, 6G (1999)\\
Jones, F.C, Ellison, D.C Sp.Sc.R 58, 259-346(1991)\\
Kirk,J.G., \& Schneider, P., ApJ 315, 425 (1987a) \\
Kirk, J.G, $\&$ Schneider, P. AJ 322, 256-265(1987)\\
Kirk,J.G., \& Webb, G.M.,  ApJ 331 , 336 (1988) \\
Lieu, R., \& Quenby, J.J., ApJ 350 , 692 (1990)\\
Meszaros,P., \& Rees, M.J.,  ApJ 405, 278 (1993)\\
Ostrowski, M., MNRAS 249, 551-559(1991)\\
Ostrowski, M. \& Schlickeiser R.,  Sol.Ph 167, 381-394(1996)\\
Peacock, J.A, MNRAS 196, 135-152(1981)\\
Quenby, J.J., \&  Lieu, R., Nature 342, 654(1989)\\
Quenby, J.J., Sp.Sc.R. 37, 201-237(1984)\\
Vietri, M., ApJ 453, 883(1995)\\
Vietri, M., Phys.Rev.Lett.78, 4328V(1997)\\
Vietri, M., ApJLet 488, L105(1998)\\
Waxman,E., Phys.Rev.Lett., 75, 386 (1995)\\

\end{document}